\documentclass[fleqn,10pt]{wlscirep}
\usepackage[utf8]{inputenc}
\usepackage[T1]{fontenc}
\usepackage{lineno}
\usepackage{comment}
\usepackage{multirow}
\usepackage{multicol}
\usepackage{xcolor}

\nolinenumbers

\title{Analyzing Reward Dynamics and Decentralization in Ethereum 2.0: An Advanced Data Engineering Workflow and Comprehensive Datasets for Proof-of-Stake Incentives}

\author[1,$\dag$]{Tao Yan}
\author[1,$\dag$]{Shengnan Li}
\author[1,$\dag$]{Benjamin Kraner}
\author[2,$\dag$,*]{Luyao Zhang}
\author[1,*]{Claudio J. Tessone}
\affil[1]{Blockchain \& Distributed Ledger Technologies Group at Department of Informatics and UZH Blockchain Center, University of Zurich, 8057 Zurich, Switzerland}
\affil[2]{Data Science Research Center and Social Science Division, Duke Kunshan University, Suzhou, 215316, China}

\affil[*]{corresponding authors: Contact Claudio J. Tessone (claudio.tessone@uzh.ch) at University of Zurich; Contact Luyao Zhang (lz183@duke.edu) at Duke Kunshan University.}

\affil[$\dag$]{these authors contributed equally to this work}

\begin{abstract}

Ethereum 2.0, as the preeminent smart contract blockchain platform, guarantees the precise execution of applications without third-party intervention. At its core, this system leverages the Proof-of-Stake (PoS) consensus mechanism, which utilizes a stochastic process to select validators for block proposal and validation, consequently rewarding them for their contributions. However, the implementation of blockchain technology often diverges from its central tenet of decentralized consensus, presenting significant analytical challenges. Our study collects consensus reward data from the Ethereum Beacon chain and conducts a comprehensive
analysis of reward distribution and evolution, categorizing them into attestation, proposer and sync committee rewards. To evaluate the degree of decentralization in PoS Ethereum, we
apply several inequality indices, including the Shannon entropy, the Gini Index, the Nakamoto Coefficient, and the Herfindahl-Hirschman Index (HHI). Our comprehensive dataset is publicly available on Harvard Dataverse, and our analytical methodologies are accessible via GitHub, promoting open-access research. Additionally, we provide insights on utilizing our data for future investigations focused on assessing, augmenting, and refining the decentralization, security, and efficiency of blockchain systems.

\end{abstract}

\begin{document}

\flushbottom
\maketitle
\thispagestyle{empty}

\section{Background \& Summary}
Blockchain technology, designed to catalyze a shift towards a decentralized and equitable digital ecosystem, sets the stage for significant advancements like Ethereum 2.0. The introduction of Ethereum 2.0 signifies a transformative development in the realm of blockchain technology, marking a departure from the established Proof-of-Work (PoW) to the Proof-of-Stake (PoS) consensus mechanism.\cite{zhang2023understand,asif2023shaping} This transition, which took place on September 15, 2022, not only aims to mitigate the environmental and scalability challenges inherent in PoW but also ushers in a novel approach to reward distribution that prioritizes staking Ether over computational exertion. Previous studies have underscored significant variations in reward distribution among different blockchain networks, such as Cardano,\cite{li2023reward} and Bitcoin,~\cite{Ashi2021Charac,karakostas2022sok} sparking debates over the potential concentration of wealth and authority.~~\cite{zhang2022sok,ao2023decentralized,zhang2023blockchain, Xiao_2023,Zhang23} Against this backdrop, the PoS iteration of Ethereum offers an invaluable opportunity to investigate whether this paradigm shift could lead to a more equitable allocation of rewards, challenging the centralization trends noted in PoW frameworks.

Notwithstanding the significance of this investigation, there is a noticeable scarcity of detailed, accessible data on reward distribution within the PoS domain of Ethereum. Our study seeks to bridge this gap by formulating a comprehensive methodology for accruing reward data from the Ethereum Beacon chain, with the goal of evaluating the degree of decentralization in this emergent ecosystem. By implementing Ethereum Erigon and Teku nodes to harvest data from the Beacon chain and employing a variety of inequality indices to scrutinize the decentralization of reward allocation, our research sheds light on the distribution patterns of rewards among validators in Ethereum 2.0, suggesting a progression towards a more decentralized reward framework.

Our paper has three main contributions, which are as follows:
\begin{enumerate}
\item We have developed a comprehensive and systematic methodology for the collection of Ethereum's reward data specific to the Proof-of-Stake (PoS) phase. This dataset, significant for its relevance and scope, has been made publicly accessible to facilitate and encourage further scholarly research.
\item Through our detailed analysis of the reward distribution in Ethereum's PoS phase, employing a range of sophisticated inequality metrics, we have identified a notable reduction in the disparity of rewards in terms of the values of inequality 
metrics. This finding is pivotal in understanding the economic dynamics of the PoS system.
\item Our dataset lends itself to a variety of analytical applications, encompassing not only time-series analysis but also an exploration of inter-layer blockchain decentralization. Furthermore, it provides a foundation for a comparative analysis of reward distributions between PoS and PoW blockchain architectures, offering critical insights into the evolving landscape of blockchain technology.
\end{enumerate}

To our knowledge, this is the first study to scrutinize the decentralization in Ethereum's PoS reward distribution using inequality metrics. While previous studies have delved into the decentralization of wealth in blockchain systems like Cardano, Bitcoin, and PoW Ethereum,\cite{wu2020a} our focus is on Ethereum PoS, thus providing fresh insights into its reward dynamics.

The organization of this manuscript is methodically outlined as follows: Section~\ref{sec:method} elucidates our research methodology, encompassing the deployment of archive nodes, a detailed analysis of consensus rewards, reward dynamics after the Merge, and the application of various inequality metrics. Section~\ref{sec:records} systematically presents the data record. In Section~\ref{sec: validation}, we undertake a technical validation of our methodology, contrasting it with additional data sources for robustness. Section~\ref{sec: usage} delves into the potential applications of our datasets, highlighting their versatility and scope. Finally, Section~\ref{sec:code} details the accessibility of our open-source code, underlining our commitment to transparency and collaborative research.
\section{Methods}
\label{sec:method}
\begin{figure}[!htbp]
    \centering
    \includegraphics[width=0.8\textwidth]{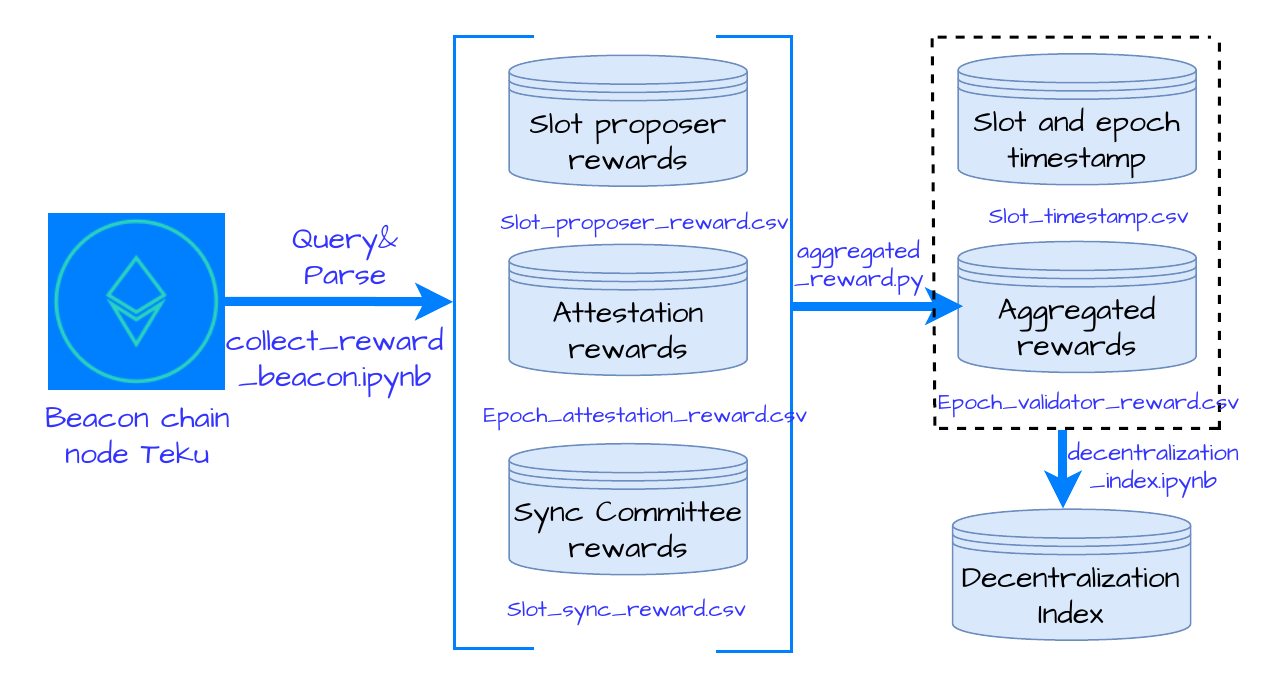}
    \setlength{\abovecaptionskip}{0.1cm}
    \caption{The Data Engineering Workflow for Beacon Chain Rewards} 
    \label{fig:data_workflow}
\end{figure}

In this section, we present the data engineering workflow, exemplified in Figure \ref{fig:data_workflow}.

\subsection*{Deployment of Archive Node for Blockchain Data Collection}

To acquire data for Ethereum 2.0, we deployed consensus and execution nodes, utilizing the Teku\footnote{\url{https://github.com/Consensys/teku}} and Erigon\footnote{\url{https://github.com/ledgerwatch/erigon}} clients\footnote{For information on Ethereum blockchain nodes and clients, please refer to \url{https://ethereum.org/en/developers/docs/nodes-and-clients/}} respectively on the Linux server, we show the experiment setting in Table \ref{tab:computer-config}. Subsequently, after synchronizing the block data, we employed the \textit{Web3.py} Python library and the API of the Teku node to collect reward data. The rewards are categorized into three main types: proposer reward, attestation reward, and sync committee reward. Notably, the attestation reward is updated per epoch. Each epoch can encompass hundreds of thousands of validators, leading to a substantial volume of data. We collected data for a duration of two months after the Ethereum PoS transition, resulting in a total dataset size of 173.8 GB, with attestation reward data constituting 167 GB. All the rewards data obtained are in Gwei, which is equivalent to \(10^{-9}\) Ether, we convert the rewards from Gwei to Ether in our analysis.

\begin{table}
\centering
    \begin{tabular}{|p{6cm}|p{6cm}|}
        \hline
        \textbf{Component} & \textbf{Specification} \\
        \hline
        CPU & 128 cores \\
        \hline
        Memory & 512 GB \\
        \hline
        Storage & 10 TB \\
        \hline
        Operating System & Debian 5.10.162-1 \\
        \hline
    \end{tabular}
    \caption{Experiment setting}
    \label{tab:computer-config}
\end{table}

\subsection*{Decomposition of Validator and Beacon Rewards by Sources}\label{sec:reward_Decomposition}

The \textit{Ethereum 2.0 protocol} incorporates a multifaceted \textit{reward system} designed to incentivize validators across various dimensions of system participation and security. Notably, these rewards can be categorized into two primary sources: those originating from the \textit{consensus layer} (Beacon chain) and those from the \textit{execution layer}. Within the Beacon chain, rewards are issued to validators in recognition of their pivotal role in the consensus mechanism, which stands as a cornerstone for the network's security and resilience. In contrast, the execution layer introduces two distinct types of rewards:

\begin{enumerate}
    \item \textbf{Gas Fees Accumulation}: This component involves the accumulation of \textit{gas fees}, which users pay to facilitate the inclusion of their transactions within a block.~\cite{liu2022empirical}
    
    \item \textbf{Maximum Extractable Value (MEV) Extraction}: MEV is associated with the value that validators can extract from the ordering and inclusion of transactions within newly created blocks.~\cite{fu2023ai} Validators have the option to optimize this process by delegating such responsibilities to specialized agents within the ecosystem, such as \textit{Flashbots}.~\cite{li2023demystifying}
\end{enumerate}

Our research focus revolves around a comprehensive examination of the reward mechanisms inherent to the consensus layer of the Beacon chain. Our primary objective is to illuminate its pivotal role in nurturing active participation and ensuring the long-term stability of the Ethereum network. It is of utmost importance to underscore that the rewards disbursed at the Beacon chain layer significantly contribute to the expansion of the monetary supply, denominated in Ether, within the system.

Validators, entities holding stakes of 32 Ether within the Beacon chain contract\footnote{https://etherscan.io/address/0x00000000219ab540356cbb839cbe05303d7705fa}, receive rewards for engaging in three distinct roles within the Proof of Stake (PoS) consensus process: \textit{attestors}, \textit{block proposers}, and members of the \textit{sync committee}:

\begin{enumerate}
    \item \textbf{Attestor} are entitled to rewards for attesting to the following:\footnote{Refer to: \url{https://eth2book.info/capella/}}
    
    \begin{enumerate}
        \item \textit{Source}: Voting in favor of a source checkpoint for Casper FFG.
        \item \textit{Target}: Voting in favor of a target checkpoint for Casper FFG.
        \item \textit{Head}: Voting for a chain head block for LMD-GHOST.
    \end{enumerate}
    
    \item \textbf{Block Proposers} receive rewards in three different categories:
    
    \begin{enumerate}
        \item \textit{Attestation}: Inclusion of attestations in a Beacon chain block.
        \item \textit{Sync Committee}: Incorporation of the sync committee's output.
        \item \textit{Whistleblowing}: Reporting instances of malicious behavior, which encompasses:
        
        \begin{enumerate}
            \item \textit{Proposer Slashing}: Reporting a slashable violation by a proposer.
            \item \textit{Attestor Slashing}: Reporting a slashable violation by an attestor.
        \end{enumerate}
    \end{enumerate}
    
    \item \textbf{Sync Committee Members} play a pivotal role in assisting light clients in maintaining a synchronized record of Beacon block headers.
\end{enumerate}

This categorization establishes the groundwork for an in-depth analysis of Ethereum 2.0's reward distribution mechanisms within its Beacon chain consensus layer.

\subsection*{Reward dynamics after the Merge}
\begin{figure}[ht]
    \centering
    \includegraphics[width=0.6\textwidth]{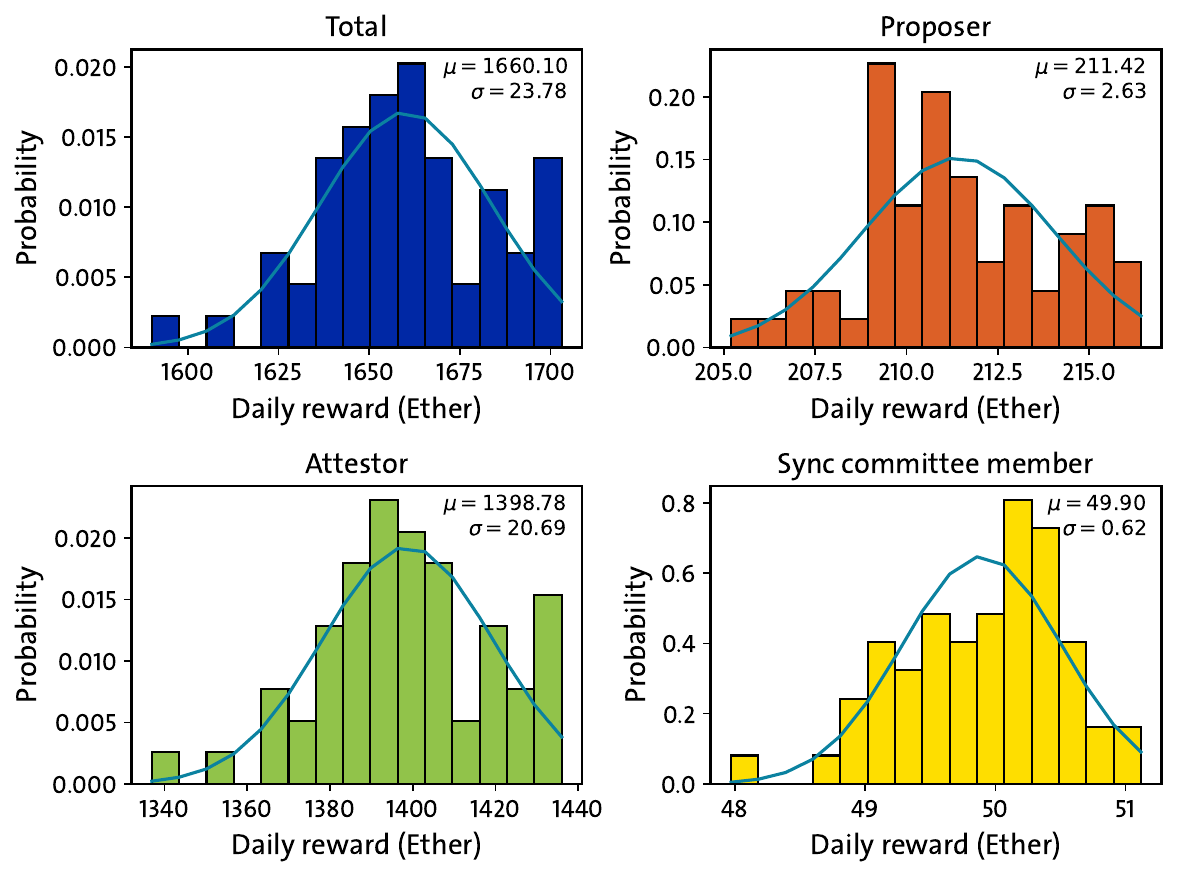}
    \setlength{\abovecaptionskip}{0.1cm}
    \caption{Distributions of the daily total, attestation, proposer, and sync committee rewards} 
    \label{fig:daily_distribution_reward}
\end{figure}

\begin{figure}[ht]
    \centering
    \includegraphics[width=0.6\textwidth]{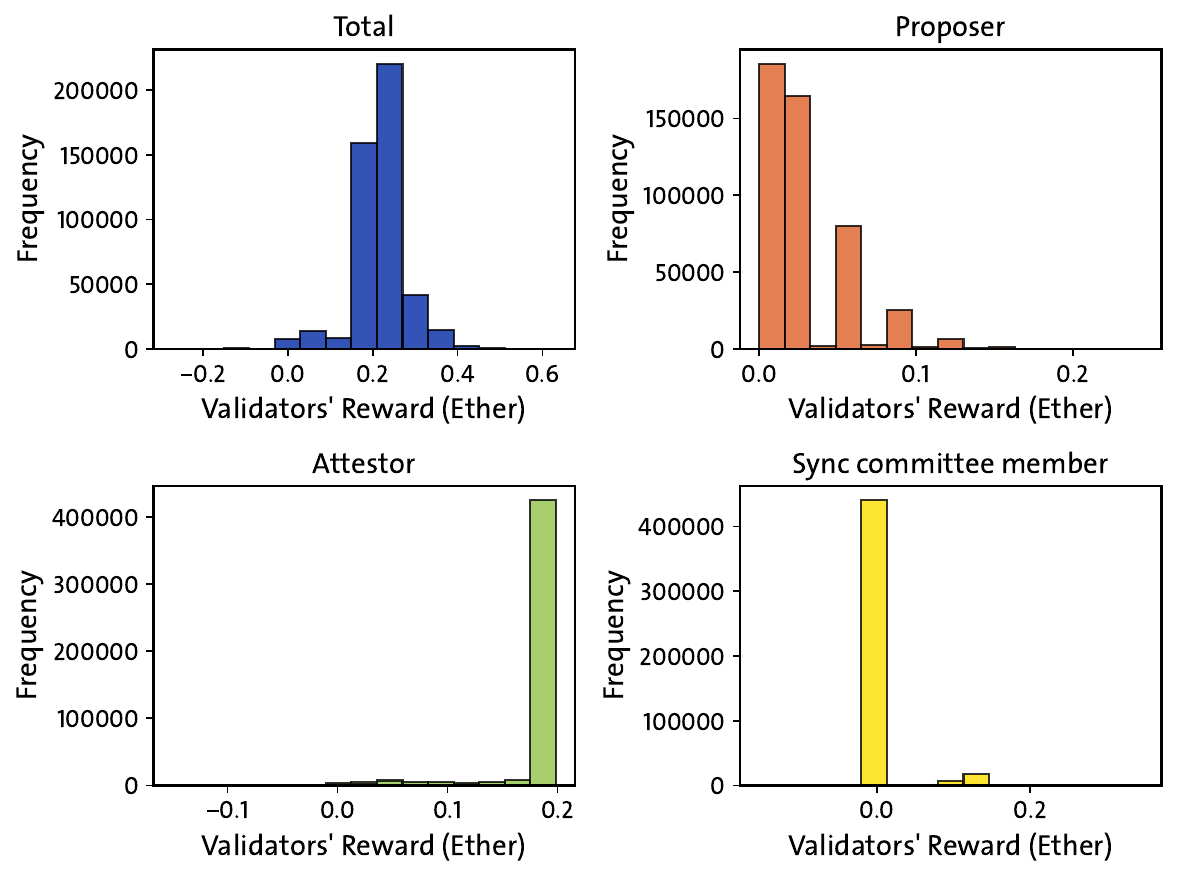}
    \setlength{\abovecaptionskip}{0.1cm}
    \caption{Distributions of the total, attestation, proposer, and sync committee reward among validators} 
    \label{fig:validator_distribution_reward}
\end{figure}

\begin{figure}[!htbp]
    \centering
    \includegraphics[width=0.8\textwidth]{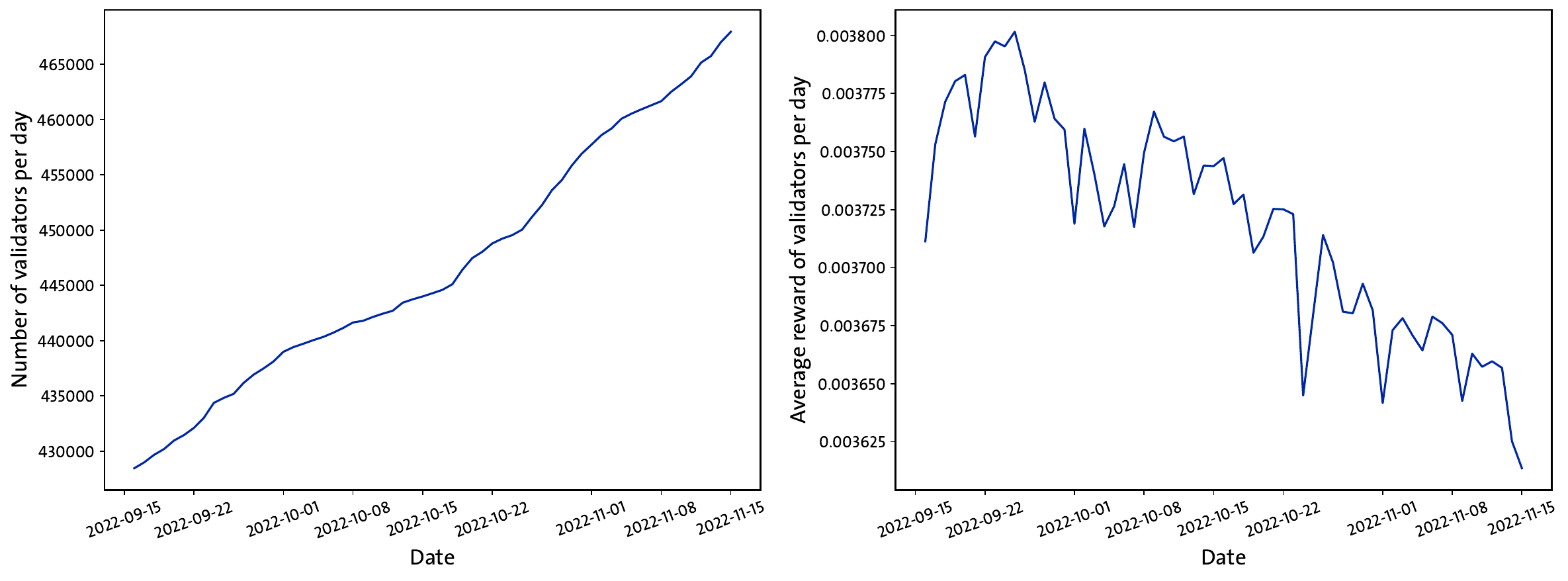}
    \setlength{\abovecaptionskip}{0.1cm}
    \caption{Growth of validators and the average reward of validators} 
    \label{fig: validator_reward_mean}
\end{figure}

\begin{enumerate}
\item \textbf{Daily reward evolution}  Figure~\ref{fig: reward3_timeseries_daily} illustrates a time series of the rewards from September 16 to November 15, 2022. The Figure shows the daily aggregated rewards on the Beacon chain, categorized into three distinct types as defined in Section~\ref{sec:method}. The blue line in Figure~\ref{fig: reward3_timeseries_daily} represents the total daily rewards issued on the Beacon chain. Our analysis reveals a consistent trend in the daily issuance of rewards, with a slight upward trajectory over the period. Furthermore, Figure~\ref{fig: reward3_timeseries_daily} offers insights into the allocation of rewards to validators for their duties. The majority of the rewards are attributed to attestation duties, which are the most frequent tasks performed once per epoch by each validator. This is followed by a significantly smaller portion allocated to proposer duties, and the smallest share comes from the participation in sync committees. 
\begin{figure}[!htbp]
    \centering
    \includegraphics[width=0.8\textwidth]{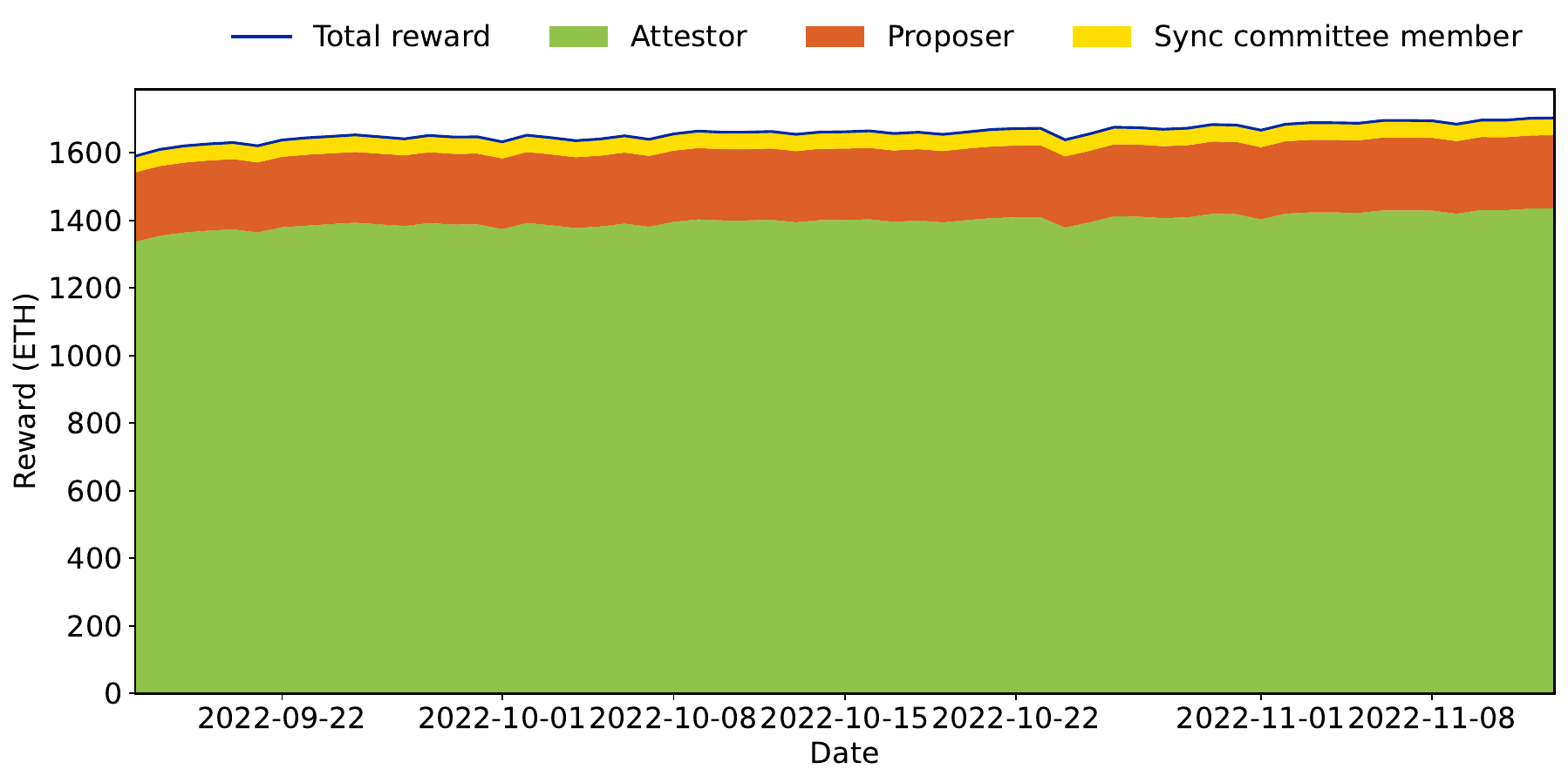}
    \setlength{\abovecaptionskip}{0.1cm}
    \caption{Daily Attestation, proposer, and sync committee reward} 
    \label{fig: reward3_timeseries_daily}
\end{figure}

 \item \textbf{Reward Distribution}
 Figure~\ref{fig:daily_distribution_reward} shows histograms of the (daily) aggregated rewards, split into the total reward and the three categories. Through four types of reward distribution, we can obtain fundamental knowledge about Beacon rewards. On average, 1660.1 Ether are generated daily on the Beacon chain. Out of this, 211.42 Ether are allocated to proposers, 1398.78 Ether are allocated to attestors, and 49.9 Ether are allocated to Sync committee members. Furthermore, the distributions expose a low dispersion,
which is comparable among each of them. A different picture emerges, when looking at the single validators operating on the Beacon chain. Figure~\ref{fig:validator_distribution_reward} shows the histograms of the income of validators (aggregated over the whole study period). On average, a validator can earn 0.25 Ether per day. A validator is frequently selected as an attestor, and can consistently earn around 0.2 Ether per day. If a validator is chosen as a proposer or a sync committee member, they will receive more rewards. The total
reward is centered around a mean value, exhibiting a very low dispersion. This indicates that on a
validator level rewards are distributed fairly. This distribution is an indication, that the protocol
works as intended, no larger group of validators is able to extract substantially more rewards. Negative values are explained by validator being ”slashed” for not conforming behavior. 
Finally, the bottom right panel of Figure~\ref{fig:validator_distribution_reward} shows the rewards validators get from being a member on a Sync committee. As the probability of being part of a sync committee is very low, we can see that a large number of validators did not receive any rewards in this category.

\item \textbf{Validators Growth}
The left panel of Figure~\ref{fig: validator_reward_mean} shows how the number of active validators evolved during a period of two months. It steadily increased by around 5 percent during the whole period. The average reward of a single validator, as shown in the right panel of Figure~\ref{fig: validator_reward_mean}, decreased slightly. The decrease in the average daily reward for a validator can be attributed to the slower pace of Ether issuance on the Beacon chain compared to the rate of validator growth. As additional validators are incorporated into the network, the mean reward allocated to each validator is expected to decline progressively.

\end{enumerate}


\subsection*{Application of Inequality Metrics to Evaluate Decentralization}

Asahi et al.(2021) \cite{Ashi2021Charac} conducted an investigation into the distribution of wealth in eight major cryptocurrencies, including Bitcoin and the Proof-of-Work (PoW) Ethereum. Their findings revealed that, despite the purported emphasis on decentralization within various blockchain networks, wealth distribution remained unequal, with the notable exception of Dash coin. To assess the decentralization of Ethereum 2.0, we employ several decentralization metrics as outlined in "SoK: Blockchain Decentralization~\cite{zhang2022sok}." These metrics offer valuable insights into the concentration of rewards among different stakeholders. Definitions of the metrics are provided in Section~\ref{appendix:metric}. 

\begin{figure}[!htbp]
    \centering
    \includegraphics[width=0.8\textwidth]{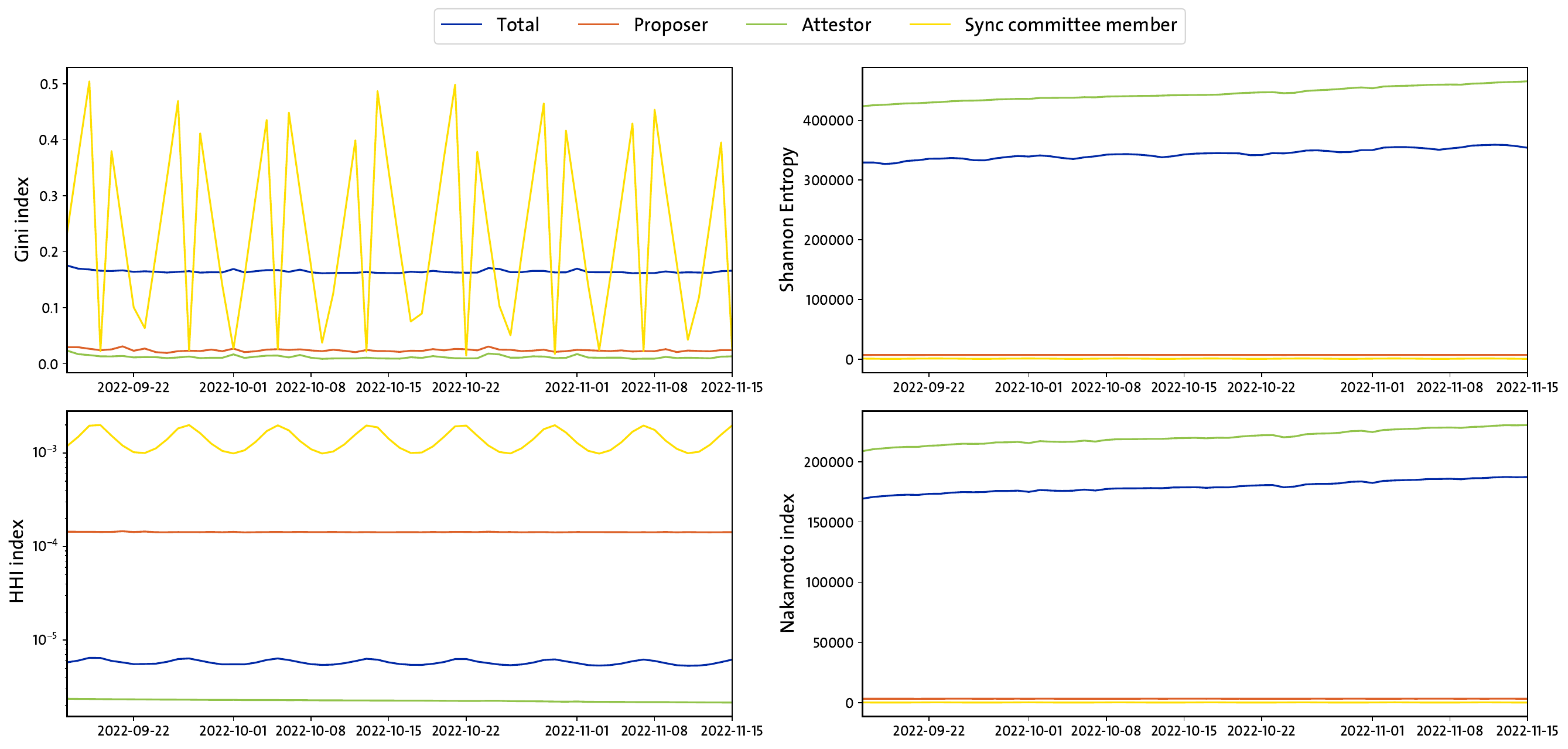}
    \setlength{\abovecaptionskip}{0.1cm}
     \caption{Inequality indices: The Gini index (top left), Shannon Entropy (top right), HHI (bottom left) and the Nakamoto index (bottom right) accumulated and split into the single reward bearing categories. All these metrics measure the reward inequality among validators who have received corresponding rewards.}
    \label{fig: index_timeseries}
\end{figure}

The first main observation from Figure~\ref{fig: index_timeseries} is that the distribution of total reward, attestor reward, and proposer reward in Ethereum 2.0 is relatively equal, as indicated by the Gini index being below 0.2 and the HHI index being very small. Additionally, the Shannon Entropy and Nakamoto index show high levels of diversity in rewards. A second crucial finding from the plot is that all metrics remain relatively stable over the chosen period, indicating a consistent level of decentralization over time. Moreover, there is a slight trend towards greater decentralization. The stability of these indices is a positive aspect, as it reaffirms that the design of Ethereum's PoS mechanism effectively prevents validators from becoming centralized at the validator level. Lastly, the reward distribution among sync committee members is neither equal nor stable, exhibiting periodicity. This discrepancy can be attributed to the mechanism described in Section~\ref{sec:reward_Decomposition} that the sync committee is only assigned every 256 epochs and comprises 512 validators. 

\section{Data Records}
\label{sec:records}

\subsection{Slot and epoch timestamp}\label{Slot and epoch timestamp}
When querying reward data from the Beacon chain, slot number or epoch number is returned without corresponding timestamps. To gain a more comprehensive understanding of the dynamics of rewards, it is imperative to synchronize the timestamps with each slot and epoch. Fortunately, the time gap between slots and epochs on the Beacon chain is fixed. The smallest time unit on the Beacon chain is a slot, with each slot lasting 12 seconds and every 32 slots forming an epoch. By knowing the start time of the first slot which is 1606824023\footnote{https://beaconcha.in/slot/0}, the time of each subsequent slot can be deduced. As for the time of each epoch, we take the time of the first slot within each epoch as the epoch's time. The data structure of slot and epoch timestamp is shown in Table ~\ref{tab:timestamp}.
\begin{table}[h]
\centering
\begin{tabular}{|p{6cm}|p{6cm}|}
\hline
\textbf{Variable} & \textbf{Data Type} \\ 
\hline
slot         & int64 \\ 
\hline
epoch        & int64 \\ 
\hline
timestamp    & int64 \\ 
\hline
\end{tabular}
\caption{Metadata of the slot and epoch timestamp data}
\label{tab:timestamp}
\end{table}
\subsection{Proposer, attestation and Sync committee reward reward}
The proposer reward data is collected from the Teku node on a per-slot basis. This data consists of various fields, namely: \texttt{proposer\_index}, \texttt{total}, \texttt{attestations}, \texttt{sync\_aggregate}, \texttt{proposer\_slashings}, \texttt{attester\_slashings}, \texttt{slot}, and \texttt{epoch}. The \texttt{total} field represents the sum of other reward types, while the \texttt{epoch} is generated based on the slot number referred to in section \ref{Slot and epoch timestamp}. Similarly, the sync committee reward is obtained on a per-slot basis and comprises the fields: \texttt{validator\_index}, \texttt{reward}, \texttt{slot}, and \texttt{epoch}. The attestation reward is acquired per epoch and includes the fields: \texttt{validator\_index}, \texttt{head}, \texttt{target}, \texttt{source}, \texttt{total\_attestation\_reward} and \texttt{epoch}. It is worth noting that all rewards are denominated in Gwei, which is equivalent to \(10^{-9}\) Ether. Thus, for our analysis, we convert the rewards from Gwei to Ether. Furthermore, validators may face penalties, resulting in negative rewards, if they fail to fulfill their duties such as missing a slot or providing invalid proposals or attestations. The structure of each reward data type is presented in Table~\ref{table:comprehensive_metadata of three rewards}.
\begin{table}[!hb]
\centering
\begin{tabular}{|c|c|c|c|c|c|c|c|}
\hline
\multicolumn{3}{|c|}{\textbf{\makebox[3cm][c]{File Name}}} & \multicolumn{1}{c|}{\textbf{\makebox[2cm][c]{Variable}}} & \multicolumn{2}{c|}{\textbf{\makebox[3cm][c]{Data Type}}} & \multicolumn{2}{c|}{\textbf{\makebox[2cm][c]{Unit}}} \\ \hline
\multicolumn{3}{|c|}{\multirow{8}{*}{\textbf{proposer reward}}} & validator\_index & \multicolumn{2}{c|}{int64} & \multicolumn{2}{c|}{count} \\ \cline{4-8} 
\multicolumn{3}{|c|}{} & total & \multicolumn{2}{c|}{int64} & \multicolumn{2}{c|}{Ether} \\ \cline{4-8} 
\multicolumn{3}{|c|}{} & attestations & \multicolumn{2}{c|}{int64} & \multicolumn{2}{c|}{Ether} \\ \cline{4-8} 
\multicolumn{3}{|c|}{} & sync\_aggregate & \multicolumn{2}{c|}{int64} & \multicolumn{2}{c|}{Ether} \\ \cline{4-8} 
\multicolumn{3}{|c|}{} & proposer\_slashings & \multicolumn{2}{c|}{int64} & \multicolumn{2}{c|}{Ether} \\ \cline{4-8} 
\multicolumn{3}{|c|}{} & attester\_slashings & \multicolumn{2}{c|}{int64} & \multicolumn{2}{c|}{Ether} \\ \cline{4-8} 
\multicolumn{3}{|c|}{} & slot & \multicolumn{2}{c|}{int64} & \multicolumn{2}{c|}{count} \\ \cline{4-8} 
\multicolumn{3}{|c|}{} & epoch & \multicolumn{2}{c|}{int64} & \multicolumn{2}{c|}{count} \\ \hline

\multicolumn{3}{|c|}{\multirow{4}{*}{\textbf{sync committee reward}}} & validator\_index & \multicolumn{2}{c|}{int64} & \multicolumn{2}{c|}{count} \\ \cline{4-8} 
\multicolumn{3}{|c|}{} & sync\_reward & \multicolumn{2}{c|}{int64} & \multicolumn{2}{c|}{Ether} \\ \cline{4-8} 
\multicolumn{3}{|c|}{} & slot & \multicolumn{2}{c|}{int64} & \multicolumn{2}{c|}{count} \\ \cline{4-8} 
\multicolumn{3}{|c|}{} & epoch & \multicolumn{2}{c|}{int64} & \multicolumn{2}{c|}{count} \\ \hline

\multicolumn{3}{|c|}{\multirow{6}{*}{\textbf{attestation reward}}} & validator\_index & \multicolumn{2}{c|}{int64} & \multicolumn{2}{c|}{count} \\ \cline{4-8} 
\multicolumn{3}{|c|}{} & head & \multicolumn{2}{c|}{int64} & \multicolumn{2}{c|}{Ether} \\ \cline{4-8} 
\multicolumn{3}{|c|}{} & target & \multicolumn{2}{c|}{int64} & \multicolumn{2}{c|}{Ether} \\ \cline{4-8} 
\multicolumn{3}{|c|}{} & source & \multicolumn{2}{c|}{int64} & \multicolumn{2}{c|}{Ether} \\ \cline{4-8} 
\multicolumn{3}{|c|}{} & total\_attestation\_reward & \multicolumn{2}{c|}{int64} & \multicolumn{2}{c|}{Ether} \\ \cline{4-8} 
\multicolumn{3}{|c|}{} & epoch & \multicolumn{2}{c|}{int64} & \multicolumn{2}{c|}{count} \\ \hline

\end{tabular}
\caption{Comprehensive Metadata for Data Files. This Table provides an organized overview of metadata for various data files related to the proposer, attestation and sync committee reward, detailing variables, data types, unit.}
\label{table:comprehensive_metadata of three rewards}
\end{table}

\subsection{Total rewards}
After obtaining the proposer reward, attestation reward and sync committee reward, we assimilate these three types of rewards into a single table called "total rewards," using the epoch number and validator index as key identifiers. This data set shows the rewards received of each validator who participates validation process of the consensus. This dataset is significantly large, measuring 174G in size, to process and analyze the data effectively, we employ the epoch timestamp outlined in section ~\ref{Slot and epoch timestamp}. By assigning a timestamp to each epoch, we are able to categorize the dataset on a daily basis. Consequently, we generate the \texttt{total\_reward\_by\_date} table, which is 1.3 G in size. This table provides information on each validator's proposer reward, attestation reward, and sync committee member reward. To provide a detailed overview of the aggregated rewards, the comprehensive metadata fields pertaining to rewards by epoch and rewards by date are presented in Table ~\ref{table:comprehensive_metadata of total rewards}. 

\begin{table}[!htbp]
\centering
\begin{tabular}{|c|c|c|c|c|c|c|c|}
\hline
\multicolumn{2}{|c|}{\textbf{\makebox[3cm][c]{File Name}}} & \multicolumn{2}{c|}{\textbf{\makebox[2cm][c]{Variable}}} & \multicolumn{2}{c|}{\textbf{\makebox[3cm][c]{Data Type}}} & \multicolumn{2}{c|}{\textbf{\makebox[2cm][c]{Unit}}} \\ \hline
\multicolumn{2}{|c|}{\multirow{6}{*}{total reward by epoch}} & \multicolumn{2}{c|}{validator\_index} & \multicolumn{2}{c|}{int64} & \multicolumn{2}{c|}{count} \\ \cline{3-8} 
\multicolumn{2}{|c|}{}                                  & \multicolumn{2}{c|}{total reward by epoch}           & \multicolumn{2}{c|}{int64} & \multicolumn{2}{c|}{Ether} \\ \cline{3-8} 
\multicolumn{2}{|c|}{}                                  & \multicolumn{2}{c|}{attestation reward}    & \multicolumn{2}{c|}{int64} & \multicolumn{2}{c|}{Ether} \\ \cline{3-8} 
\multicolumn{2}{|c|}{}                                  & \multicolumn{2}{c|}{sync committee reward} & \multicolumn{2}{c|}{int64} & \multicolumn{2}{c|}{Ether} \\ \cline{3-8} 
\multicolumn{2}{|c|}{}                                  & \multicolumn{2}{c|}{proposer reward}           & \multicolumn{2}{c|}{int64} & \multicolumn{2}{c|}{Ether} \\ \cline{3-8} 
\multicolumn{2}{|c|}{}                                  & \multicolumn{2}{c|}{epoch}          & \multicolumn{2}{c|}{int64} & \multicolumn{2}{c|}{count} \\ \hline
\multicolumn{2}{|c|}{\multirow{6}{*}{total reward by date}} & \multicolumn{2}{c|}{validator\_index} & \multicolumn{2}{c|}{int64} & \multicolumn{2}{c|}{count} \\ \cline{3-8} 
\multicolumn{2}{|c|}{}                                  & \multicolumn{2}{c|}{total reward by date}           & \multicolumn{2}{c|}{int64} & \multicolumn{2}{c|}{Ether} \\ \cline{3-8} 
\multicolumn{2}{|c|}{}                                  & \multicolumn{2}{c|}{attestation reward}    & \multicolumn{2}{c|}{int64} & \multicolumn{2}{c|}{Ether} \\ \cline{3-8} 
\multicolumn{2}{|c|}{}                                  & \multicolumn{2}{c|}{sync committee reward} & \multicolumn{2}{c|}{int64} & \multicolumn{2}{c|}{Ether} \\ \cline{3-8} 
\multicolumn{2}{|c|}{}                                  & \multicolumn{2}{c|}{proposer reward}           & \multicolumn{2}{c|}{int64} & \multicolumn{2}{c|}{Ether} \\ \cline{3-8} 
\multicolumn{2}{|c|}{}                                  & \multicolumn{2}{c|}{date}          & \multicolumn{2}{c|}{date} & \multicolumn{2}{c|}{count} \\ \hline

\end{tabular}
\caption{Comprehensive Metadata for total reward.}
\label{table:comprehensive_metadata of total rewards}
\end{table}

The comprehensive dataset detailing final reward records for Ethereum validators is securely stored and publicly accessible on the Harvard Dataverse.\cite{DVN/OKQRS1_2024} This dataset encompasses validator rewards from three distinct sources, presented at various frequencies, along with daily decentralization indices, all formatted in CSV. 
\section{Technical Validation}
\label{sec: validation}
To ensure the accuracy of the data, two methods of verification are employed. The first method involves crosschecking the total reward data with the statistics charts available on Beaconscan. Specifically, the "Total Daily Income (Ether)" chart\footnote{\url{https://beaconscan.com/stat/validatortotaldailyincome}} provides insights into the total rewards received by validators in Ether on a daily basis. This data can be compared with the daily total rewards highlighted in our study to ascertain their consistency.

Furthermore, for detailed verification of rewards allocated to a specific validator for a given slot or epoch number, the "Income detail history" API method\footnote{\url{https://beaconcha.in/api/v1/docs/index.html\#/Validator/get_api_v1_validator__indexOrPubkey__incomedetailhistory}} proves useful. By entering the validator's index and the epoch number on this webpage, it is possible to retrieve a breakdown of income components for that validator during the specified epoch. It is pertinent to note that the total incomes depicted include not only beacon chain income but also transaction fees from the execution layer and MEV (Maximal Extractable Value) rewards. In Table~\ref{tab:Income}, various income types are listed. For the purposes of this verification, we focus on the beacon chain incomes – attestation reward, proposer reward, and sync reward at epoch 209985 – for validator 480908, aligning with the data we have examined.

\begin{table}[!htbp]
\centering
\definecolor{lightblue}{rgb}{0.68, 0.85, 0.9}
\definecolor{lightpink}{rgb}{1.0, 0.71, 0.76}
\begin{tabular}{|l|l|l|}
\hline
\textbf{Income Type}                           & \textbf{Amount}            & \textbf{Description} \\ \hline
Attestation Source Reward                      & 3151                       & \multirow{3}{*}{Attestation reward in gwei} \\ \cline{1-2}
Attestation Target Reward                      & 5852                       &  \\ \cline{1-2}
Attestation Head Reward                        & 3132                       &  \\ \hline
Proposer Attestation Inclusion Reward          & 35844119                   & Proposer reward \\ \hline
Proposer Sync Inclusion Reward                 & 1243368                    & Sync reward \\ \hline
Transaction Fee Reward (Wei)                   & 32151870287000820          &  \\ \hline
Epoch                                          & 209985                     & Key for searching reward by epoch \\ \hline
Validator Index                                & 480908                     & Key for searching reward by validator \\ \hline
\end{tabular}
\caption{Data Validation. This table enumerates the various types of income received by Ethereum validators, highlighting the specific amounts and categorizing them into attestation, proposer, and sync rewards, along with key identifiers for data validation purposes.}
\label{tab:Income}
\end{table}
\section{Usage Notes}
\label{sec: usage}
\subsection{Applicability}
The reward dataset provided by this study enables a broad spectrum of subsequent analyses and applications. The following usage notes delineate several prospective research trajectories:

\begin{enumerate}
\item \textbf{Time Series Analysis of Rewards}: Our dataset facilitates a thorough temporal analysis of reward distribution in Ethereum's Proof of Stake (PoS) mechanism and further supports the application of machine learning techniques to blockchain data.~\cite{zhang2023machine} This analysis is crucial for elucidating temporal patterns, detecting fluctuations over time, and identifying shifts in the decentralization dynamics of the Ethereum PoS system.~\cite{saad2021pos}

\item \textbf{Inter-Layer Blockchain Decentralization Correlations}: By establishing a decentralization index at the consensus layer of the Ethereum blockchain, our dataset lays the groundwork for probing the correlations between decentralization at the consensus level and other layers of blockchain architecture, such as hardware, data, network, and application layers~\cite{chemaya2023uniswap}. This approach aligns with the future research directions suggested by Zhang (2023)~\cite{zhang2022sok} and offers a nuanced understanding of how decentralization metrics at various strata interact and exert mutual influence.

\item \textbf{Comparative Analysis of Reward Distribution in PoS vs PoW}: Utilizing this dataset to conduct a comparative analysis of reward distributions in Ethereum's Proof of Stake (PoS) and Proof of Work (PoW) stages provides vital insights. It facilitates an exploration similar to the causal inference study on transaction fee mechanisms around Ethereum's transition from PoW to PoS.~\cite{zhang2023understand,kapengut2023event} Such comparative studies are instrumental in discerning the impact of this transition on the ecosystem, particularly concerning aspects of decentralization and distributional fairness.
\end{enumerate}

\subsection{Future Research}

While our daily indices provide a comprehensive foundation, we anticipate that future research can further enhance and expand upon them. Potential avenues for future work include:

\begin{itemize}

\item \textbf{Decentralization analysis at entity and staking pool levels}: Our current work analyzed decentralization at the validator level, but the reward data could also be examined at both individual entity and staking pool levels.~\cite{bahrani2024centralization,gersbach2022staking,tang2023pool,he2020staking} Comparing and contrasting the reward dynamics between entities and pools could reveal insights about distribution differences across various levels.

\item \textbf{Connecting Beacon Rewards to Block Building Rewards}: As shown in Table~\ref{tab:Income}, validators receive transaction fee rewards~\cite{leonardos2021dynamical} for participating in block building, in addition to beacon rewards. Moreover, validators also gain other miner extractable value(MEV)~\cite{mancino2023exploiting} from the block-building process. Future efforts could incorporate analyses of these additional validator revenue sources.

\item \textbf{Connecting Blockchain to Real-World Assets}: Our current dataset measures rewards in Ethereum's native currency. Further research could account for Ethereum's inflation or deflation over time,~\cite{conlon2021inflation} as well as exchange rates between Ethereum and other crypto or fiat currencies, usually studied in financial technology~\cite{zhu2023educational,yu2023bitcoin,Zhang_2023,9881799,10.1007/978-3-031-37717-4_51} and blockchain interoperability studies~\cite{augusto2023sok,belchior2021survey}. This would connect on-chain decentralization to real-world asset values.

\end{itemize}

In summary, while our work establishes a robust baseline, there are many exciting opportunities to build on it through additional layers of analysis and connections to external factors. We hope our indices spur further decentralization research that encompasses new dimensions and perspectives.


\section{Code Availability}
\label{sec:code}
The datasets and Python codebase employed for the analysis of beacon chain rewards in the Ethereum 2.0 PoS (Proof of Stake) framework are available in a public repository on GitHub at \url{https://github.com/learn-want/ETH2.0-reward}. This code, primarily developed in Python and encapsulated within Jupyter Notebook environments, facilitates comprehensive investigations into the dynamics of consensus rewards on the Ethereum blockchain. Academics, blockchain developers, and other stakeholders are encouraged to leverage this open-source resource for advanced studies and explorations in blockchain reward mechanisms.


\appendix
\section{Introduction to Decentralization Metrics}
\label{appendix:metric}
\subsection{Shannon entropy}
Shannon entropy is a measure of the uncertainty or randomness in a probability distribution. It quantifies the concentration of rewards among different stakeholders by considering the probabilities of receiving rewards \cite{zhang2022sok}. The Shannon entropy formula is given by:

\begin{equation}
H(X) = -\sum_{i=1}^{n} P(x_i) \log_2 P(x_i)
\end{equation}

where $H(X)$ represents the Shannon entropy of the reward distribution and $P(x_i)$ is the probability of stakeholder $x_i$ receiving a reward. A higher entropy value indicates a more equal distribution of rewards.

\subsection{Gini Coefficient}
The Gini coefficient is a widely used metric to measure inequality in income or wealth. It measures the inequality among stakeholders by comparing the cumulative distribution of rewards with an ideal distribution \cite{zhang2022sok}. The Gini coefficient is calculated as follows:
where $G$ represents the Gini coefficient, $x_i$ and $x_j$ are the rewards received by stakeholders $i$ and $j$, $n$ is the number of stakeholders, and $\mu$ is the mean of all rewards. A Gini coefficient of 0 indicates perfect equality, while a coefficient of 1 represents maximum inequality. As there are negative rewards because some validators that do not complete their obligations and are punished by the system, we process these negative rewards are zero.
\begin{equation}
G = \frac{\sum_{i=1}^{n} \sum_{j=1}^{n} |x_i - x_j|}{2n^2 \mu}
\end{equation}

\subsection{Herfindahl-Hirschman Index (HHI)}
The Herfindahl-Hirschman Index (HHI) is another measure of market concentration often used in economics. It measures the concentration of rewards among stakeholders by summing the squares of their market shares \cite{zhang2022sok}. The HHI is calculated as follows:

\begin{equation}
HHI = \sum_{i=1}^{n} s_i^2
\end{equation}

where $HHI$ represents the Herfindahl-Hirschman Index and $s_i$ is the market share of stakeholder $i$. A higher HHI value indicates a higher concentration of rewards among a few stakeholders.

\subsection{Nakamoto Coefficient}
The Nakamoto coefficient measures the extent to which the majority of rewards are concentrated among a small number of stakeholders. It calculates the minimum percentage of stakeholders required to control more than 50\% of the rewards \cite{zhang2022sok}. The Nakamoto coefficient is defined as:

\begin{equation}
N = \frac{\sum_{i=1}^{m} r_i}{\sum_{i=1}^{n} r_i}
\end{equation}

where $N$ represents the Nakamoto coefficient, $r_i$ is the reward received by the top $m$ stakeholders, and $n$ is the total number of stakeholders. A higher Nakamoto coefficient indicates a higher degree of centralization.

\bibliography{sample}


\section*{Acknowledgements} 

Luyao Zhang is supported by National Science Foundation China on the project entitled “Trust Mechanism Design on Blockchain: An Interdisciplinary Approach of Game Theory, Reinforcement Learning, and Human-AI Interactions (Grant No. 12201266). Luyao Zhang is also with SciEcon CIC, a not-for-profit organization based in the United Kingdom, aiming to cultivate interdisciplinary research of profound insights and practical impacts.

\end{document}